\documentclass{article}

\usepackage{arxiv}

\usepackage{graphicx}
\usepackage[utf8]{inputenc}
\usepackage{subcaption}

\usepackage{xcolor}
\usepackage{relsize,amsfonts}
\newcommand\bigexists{%
  \mathop{\lower0.75ex\hbox{\ensuremath{%
    \mathlarger{\mathlarger{\mathlarger{\mathlarger{\exists}}}}}}}%
  \limits}
\title{Learning Attacker's Bounded Rationality Model in Security Games}

\author{
  Adam {\.Z}ychowski \\
  Faculty of Mathematics and Information Science, \\
  Warsaw University of Technology \\
  Warsaw, Poland\\
  \texttt{a.zychowski@mini.pw.edu.pl} \\
   \And
  Jacek Ma{\'n}dziuk \\
  Faculty of Mathematics and Information Science, \\
  Warsaw University of Technology \\
  Warsaw, Poland\\
  \texttt{j.mandziuk@mini.pw.edu.pl} \\
}

\begin{document}
\maketitle

\begin{abstract}
The paper proposes a novel neuroevolutionary method (NESG) for calculating leader’s payoff in Stackelberg Security Games. The heart of NESG is strategy evaluation neural network (SENN). SENN is able to effectively evaluate leader’s strategies against an opponent who may potentially not behave in a perfectly rational way due to certain cognitive biases or limitations. SENN is trained on historical data and does not require any direct prior knowledge regarding the follower’s target preferences, payoff distribution or bounded rationality model. 
%SENN was incorporated and tested with Evolutionary Algorithm for Security Games (EASG). 
NESG was tested on a set of 90 benchmark games inspired by real-world cybersecurity scenario known as deep packet inspections. Experimental results show an advantage of applying NESG over the existing state-of-the-art methods when playing against not perfectly rational opponents. The method provides high quality solutions with superior computation time scalability.
%vs state-of-the-art methods. 
Due to generic and knowledge-free construction of NESG, the method may be applied to various real-life security scenarios.
%that include homeland security, cybercsecurity or wildlife protection. 
%The paper discusses one of such application in cybersecurity domain, named deep packet inspections.

\end{abstract}

\section{Introduction}\label{sec:Introduction}
%Computational Intelligence solutions are widely applied to various domains, e.g. medicine,
%~\cite{becker2019artificial}, 
%transport,
%~\cite{del2019bioinspired}, 
%recommendations,
%~\cite{zhang2018explainable}, 
%or communication.
%~\cite{sailunaz2018emotion}. 
One of the salient application domains of Computational Intelligence is security management (surveillance, guards patrolling, anti-poaching operations, border controlling, cybersecurity, anti-terrorism etc.). 
%The popular approach in security management supporting systems is to put a game theory framework on a real-world scenario and model it as a game between defenders (e.g. protectors, guards, police) and attackers (e.g. thieves, poachers, terrorists). 

Within security management one of the most popular computational models of attacker-defender scenarios are Security Games (SG)~\cite{sinha2018stackelberg}.
%
%\subsection{Security Games} 
SGs follow the idea of Stackelberg games~\cite{Leitmann_1978} and involve two players: the \emph{leader} and the \emph{follower}. SG consists of two phases. First, the leader commits to a certain strategy. Next, the follower, based on the leader's decision, chooses their strategy. This sequence of decisions establishes information asymmetry (favouring the follower) and mimics real-world scenarios in which the attacker (follower) can observe the defender's (leader's) strategy (e.g. patrol schedules) and plan their attack accordingly.

The goal of SG is to find a pair of players’ strategies that form the so-called Stackelberg Equilibrium (StE)~\cite{Leitmann_1978}. 
%Since the follower is aware of the opponent’s strategy, they select a strategy that gives them the highest possible payoff. It can be realized by simple iteration over all possible follower's strategies. Thus, the problem can be reduced to finding an optimal leader's strategy which is proven to be NP-hard~\cite{conitzer2006computing}. 
For the leader, \textit{mixed strategies} are considered which are probability distributions of various deterministic strategies (a.k.a \emph{pure strategies}). 
%means that we are looking for not only \textit{pure strategy} (one strategy from a set of possible strategies) but the goal is to find the probability distribution over all pure strategies. 
Consequently, the follower despite knowing the mixed leader's strategy does not know which realization of this strategy (which pure strategy) would actually be played by the leader - he/she is only aware of the probability distribution of them. This corresponds to real-life situations, e.g. when the attacker can observe guards patrolling schedule and deducts how often a given target is visited but cannot be sure about the patrol's presence on particular time (day/hour).

Many practical applications of SGs were deployed in various security domains, including: the canine patrols scheduling system for Los Angeles International Airport,
%~\cite{jain2010software}, 
TRUSTS system for scheduling patrols for fare inspection in Los Angeles Metro,
%~\cite{yin2012trusts}, 
PROTECT system which randomizes schedules of US Coast Guard's resources in Boston harbour,
%~\cite{shieh2012protect}, 
or PAWS system to prevent poaching and protecting wildlife in Queen Elizabeth National Park in Uganda.
%~\cite{yang2014adaptive}. 
Please consult a survey paper~\cite{sinha2018stackelberg} for details.

\subsection{Bounded rationality}
One of the fundamental assumptions in SGs is that the attacker chooses an optimal response strategy. However, in real-life cases, which involve humans (e.g. terrorists, thieves, poachers, hackers), this assumption may not hold, due to limitation of the players' senses, his/her cognitive biases~\cite{kahneman2011thinking}, partial knowledge about the problem, or imprecisely defined goals~\cite{rubinstein1998modeling}.
%~\cite{aumann1997rationality,rubinstein1998modeling}. 
This deviation from optimal response selection is known as \emph{bounded rationality} (BR) behaviour~\cite{simon1957models}. Hence in practice, playing the StE strategy by the leader may be non-optimal. {\bf Considering human BR biases by the leader when selecting their strategy can potentially improve their expected results.} 
%The idea is to make a model of the world which takes into account the fact that participants' behaviour may not be perfectly rational and compute strategy according to this model. 
There are several BR models proposed in the literature, e.g.: Anchoring Theory (AT)~\cite{tversky1974judgment}, Prospect Theory (PT)~\cite{kahneman2013prospect}, Quantal Response (QR)~\cite{mckelvey1995quantal}, or Framing Effect (FE)~\cite{tversky1981framing}. While each of them is justified by psychological experiments and emphasizes different aspect of human behaviour, there is no consensus on which of them reflects the human BR bias most closely.

The vast majority of BR papers in SG domain incorporates a particular BR model (e.g. AT with $\epsilon$-optimality~\cite{cobra}, PT~\cite{yang2013improving}, QR~\cite{nguyen2013analyzing}) into single-step SG by modifying the fully-rational behavioral model
%Mixed-Integer Linear Programming (MILP) solutions. Such an approach has two main disadvantages:
which assumes possessing prior knowledge about which BR model best fits a given adversarial behaviour.
%and assumes particular BR model,
%(b) MILP are inefficient, poorly scalable and unsuitable for larger, more complicated scenarios.
{\bf In this paper, we approach the problem in a novel way by learning the actual behavioral model of the attacker based on observing his/her past performance.}
%by introducing a novel generic neural network based method that does not presuppose any specific BR model and scales visibly better than MILP based approaches.

\subsection{Motivation and Contribution}
In real-world scenarios modelled by SGs, the role of the follower is played by humans whose action selection process may be non-optimal due to certain BR biases, described above~\cite{cobra,yang2013improving,nguyen2013analyzing}. At the same time, the defenders (playing the role of the leader) usually have little or no knowledge about their adversaries (e.g. terrorists, poachers, smugglers, hackers) and have no clue about which BR model would best reflect their behaviour. Furthermore, in SG literature it is usually assumed that exact opponent's payoffs are revealed to both players, which is infeasible in real-world scenarios, where each side knows only the value of a given target to him/her, but is unaware of its precise value (financial, organizational, propagandistic, etc.) for the opponent. 
%the exact opponent's value evaluation is sometimes impossible or only approximate, for instance, stolen stuff monetization can be difficult and the price in the black market can differ from object's real value.

Due to the above reasons there is a strong need to apply generic SG solutions, that abstract from precise assumptions about the follower's BR model and his/her preferences, and are capable of inferring them from available data (e.g. analysis of the follower's past behaviour). From the leader's perspective, such a system should be able to {\bf learn} the relations between utilities and follower's decisions. Such a system is proposed in the paper.

The main contribution of the paper can be summarized as follows:
\begin{enumerate}
    \item To the best of our knowledge, this is the first successful attempt of applying neural networks to the leader's strategy estimation in SGs (SENN).
    \item Based on SENN we propose an end-to-end neuroevolutionary system (NESG) for finding high quality leader's strategies in SGs.
    \item NESG does not require any assumption about the follower's BR model or knowledge about his/her payoff distribution.
    \item Results of experiment performed in one of the cybersecurity scenarios outperform those of state-of-the-art methods in terms of both computation time and quality of results.
\end{enumerate}

\section{Problem Definition}\label{sec:Definition}

We consider $m$ step games with two players: the leader ($L$) and the follower ($F$). There is a predefined set of $n$ targets $T = \{t_1,t_2,\ldots,t_n\}$. Each target $t \in T$ is associated with 4 payoffs: $U_t^j$, $j \in \{L+, L-, F+, F-\}$ representing the leader's reward ($U_t^{L+}$), their penalty ($U_t^{L-}$), the follower's reward ($U_t^{F+}$) and their penalty ($U_t^{F-}$).

The leader possesses $k$ units. Leader's \emph{pure strategy} $\sigma^L$ is units allocation over targets in $m$ time steps. Units allocation can be different in different time steps - they can be reallocated between the steps.
Formally $\sigma^L = \{a_{us}\}$, where $a_{us} \in T$ is a target allocated for unit $u$ in time step $s$, $u \in \{1,\ldots,k\}, s \in \{1,\ldots,m\}$.

Let's denote a set of all pure strategies of the leader by $\Sigma^L$. Then, a \emph{mixed strategy} $\pi^L$ is a probability distribution over $\Sigma^L$: $\pi^L = \{(\sigma_i^L, p_i)\}$, where $p_i$ is the probability of playing strategy $\sigma_i^L \in \Sigma^L$.

\textit{Target $t$ coverage} in step $s$ (denoted by $c_s(t)$) for mixed strategy $\pi^L$ is the probability of the event that at least one unit is allocated to $t$ in step $s$ playing strategy $\pi^L$: $$c_s(t) = \sum_{\sigma_i^L \in \pi^L} p_i : \bigexists_{a_{us} \in \sigma_i^L} a_{us} = t.$$

Follower's strategy $\sigma^F$ is choosing one of the targets from $T$. Let's denote this target by $x$.
The players' payoffs are computed as follows:
\begin{itemize}
\item If in any time step leader's unit is allocated to $x$, the follower is \textit{caught} and players receive $U_x^{L^+}$ and $U_x^{F^-}$, respectively.
\item If in no time step leader's unit is allocated to $x$, the follower's attack is successful and players receive $U_x^{L^-}$ and $U_x^{F^+}$, respectively.
\end{itemize}
Therefore, expected leader's and follower's payoffs ($U^L$ and $U^F$) equal
$$U^L =  P_x U_x^{L-} + (1-P_x) U_x^{L+} \quad \textrm{and} \quad
U^F =  P_x U_x^{F+} + (1-P_x) U_x^{F-},$$
where $P_x = \prod_{s = 1,\ldots,m} 1-c_s(x)$ is probability of successful attack on target $x$.

The game model employs Stackelberg Game principles which means that first the leader commits to their strategy $\pi^L$ and then the follower, being aware of $\pi^L$, determines his/her strategy.

%czy definiowac rownowage Stackelberga, jesli jej nie uzywamy, tj. i tak nasza metoda jej nie liczy

\section{Cybersecurity scenario}\label{sec:Scenario}
Let us consider cybersecurity domain as a use case scenario. One of the methods of preventing attacks on computer networks is \textit{deep packet inspections}~\cite{el2017survey} which relies on periodical selection of a subset of packets in a computer network for analysis. 
%The procedure is costly in terms of the network's throughput and also does not check all nodes constantly. 
This problem can be formulated as SG in which the detection system plays the role of the leader and the attacker (e.g. hacker, intruder system) plays the role of the follower. Network computers (hosts) are targets. The detection system chooses a subset of hosts and inspects packets sent to them for some fixed time. Then, the next subset of hosts is checked (next time step in SG definition). If malicious packets go through undetected the attack is successful and the intruder controls the infected host. Packet inspections cause unwanted latency and the defender has to decide where to inspect network traffic in order to maximize the probability of a successful malicious packet detection. In such a scenario, the defender has no knowledge about potential invaders, their preferences or capabilities. However, historical data about attack attempts or some simulations can be used to approximate them.
\begin{figure}[ht]
	\centering
    \includegraphics[width=0.65\columnwidth]{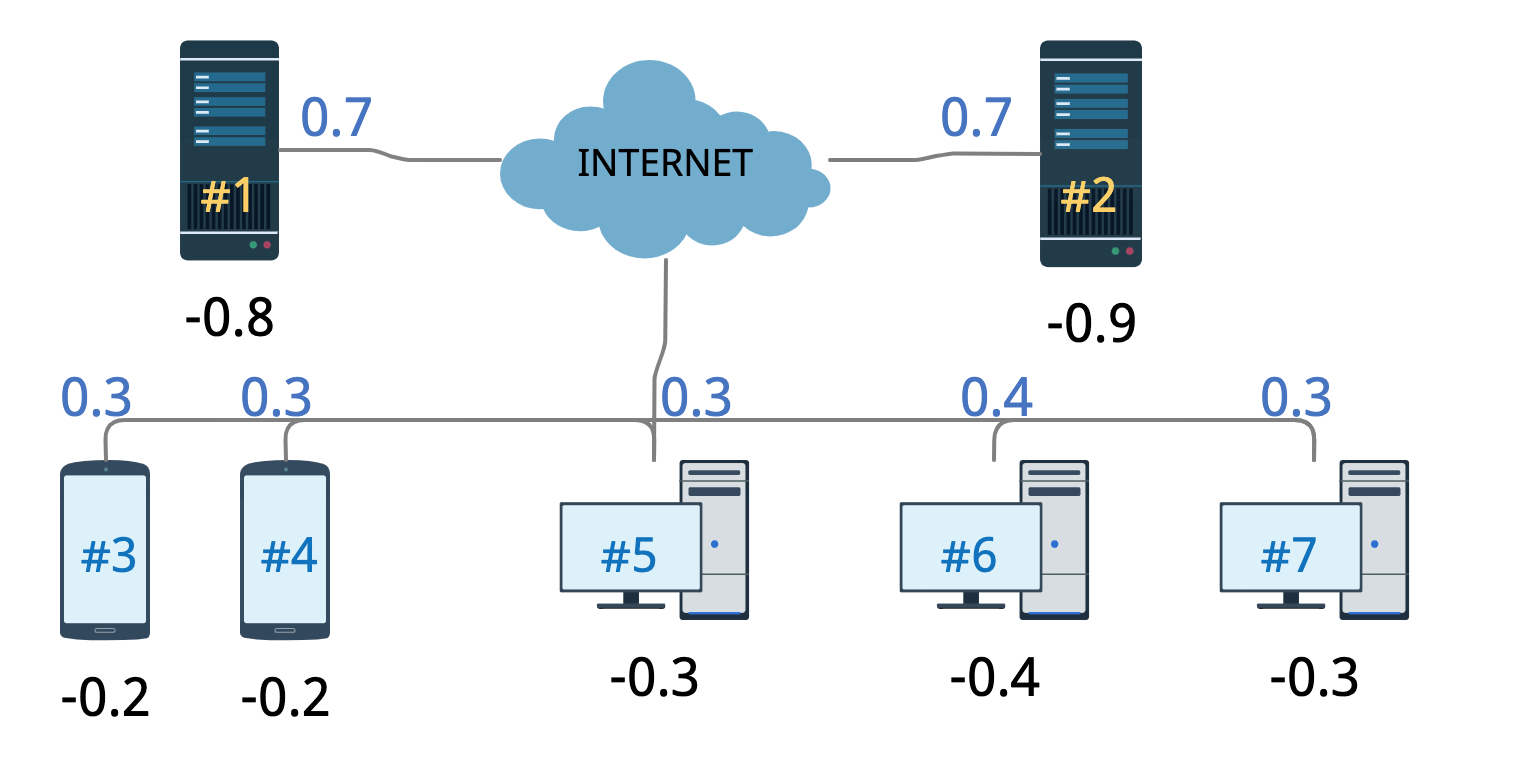}
    \caption{Sample network with 7 hosts. The numbers below hosts are leader's penalties in case of successful attack, the numbers above connections are probabilities of protecting given nodes in sample defender's strategy.}
    \label{fig:network}
\end{figure}

Figure~\ref{fig:network} shows example network with 7 hosts. For simplicity, the example refers to one-step game in which the defender does not receive any positive payoff in the case of attack detection. Defender's penalties for a successful attack are shown below each node. The network contains 2 data centres which are key infrastructure elements, 3 PCs and 2 mobile devices with the least importance. Sample defender's strategy $\pi^L$ is presented in the form of $c_s(t)$ values, which are placed above connections to particular nodes and denote the probability of protecting a given node. 
%It is targets coverage $c_s(t)$ values from game definition. 
%Sample defender's strategy corresponding with them is 
$\pi^L =\{(0.4, \{\#1,\#2,\#6\})$, $(0.3, \{\#1,\#2,\#7\})$, $(0.3, \{\#3,\#4,\#5\})\}$ and if, for instance, the attacker chooses to attack host $\#5$, the expected defender's payoff equals $-0.3(1-0.3) = -0.21$.

\section{State-of-the-art Approaches}\label{sec:SOTA}
There are two main types of SGs solution
methods: exact and approximate. 
The vast majority of exact approaches base on Mixed-Integer Linear Programming (MILP)~\cite{paruchuri2008playing},
%,jain2010software}, 
where SG is formulated as an optimization problem with linear constraints and the strategies are computed using specially optimized software engine. The main disadvantage of MILP methods is exponential time and memory scalability. The most efficient methods from this group are BC2015~\cite{bosansky2015sequence} and C2016~\cite{cermak2016using} which transform the game to equivalent form with substantially smaller MILP representation, what enables effective solving of bigger games.

Approximate methods offer a viable alternative to exact solutions and calculate close-to-optimal results much faster and for larger games, which are beyond capabilities of exact approaches. Some approximate methods are heuristic time-optimized MILP algorithms (e.g. CBK2018~\cite{vcerny2018incremental}), others employ different ideas.  O2UCT~\cite{karwowski2019stackelberg,karwowski2020AAAI}, For instance, utilizes Upper Confidence Bounds applied to trees~\cite{kocsis2006bandit} (a variant of Monte Carlo Tree Search) and combines sampling the follower's strategy space with calculating the best  leader's strategy for which a sampled followers strategy is the optimal response.
Another heuristic method, EASG~\cite{ZychowskiMandziuk2021} maintains a population of candidate leader's strategies and applies specifically designed mutation and crossover operators. EASG is designed as a general framework that can be easily adapted to various types of SG.
There are also heuristic methods devoted to particular SG formulations~\cite{nasza_AAAI}.

\section{Proposed Solution}\label{sec:Algorithm}
\begin{figure}[ht]
%	\centering
%	\begin{subfigure}[t]{0.9\linewidth}
	\centering
    \includegraphics[width=0.9\columnwidth]{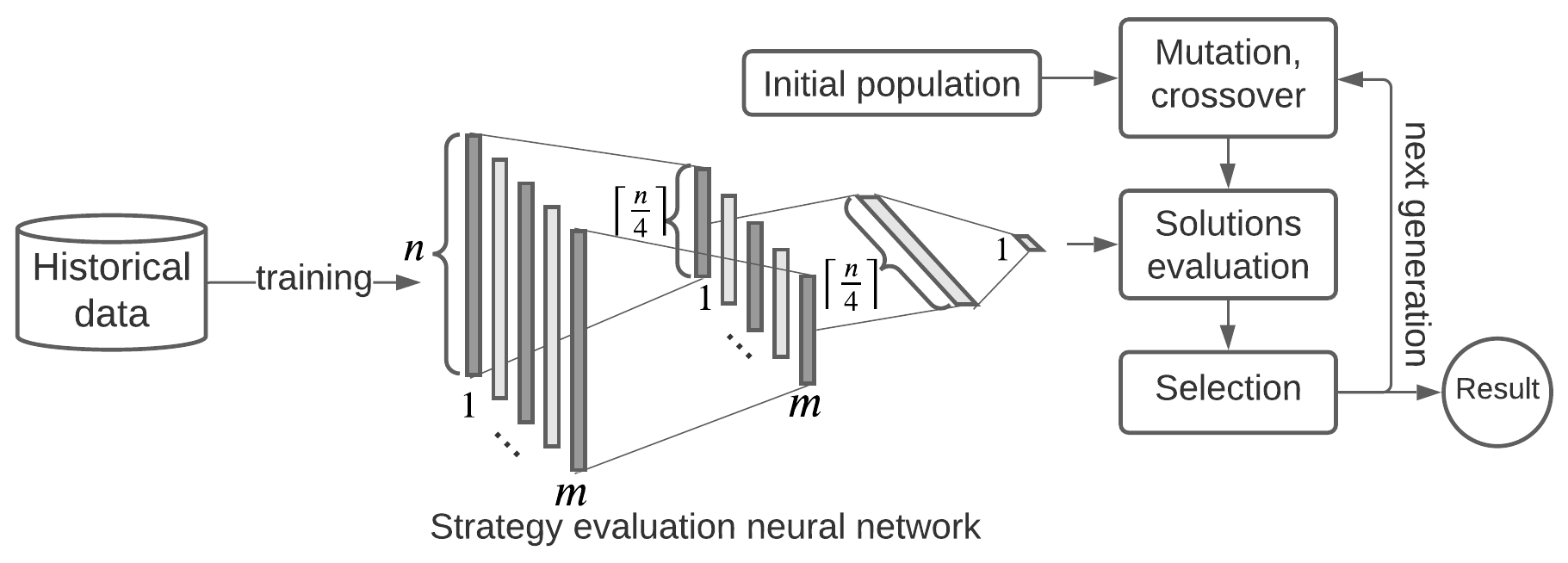}
    \caption{System overview with SENN component.}
    \label{fig:overview}
%	\end{subfigure}
%	
%    \begin{subfigure}[t]{0.7\linewidth}
%    \centering
%    \includegraphics[width=0.7\columnwidth]{images/SENN.pdf}
%    \caption{Strategy evaluation neural network architecture.}
%    \label{fig:senn}
%	\end{subfigure}
%	\caption{The proposed system components overview.}
\end{figure}

In this paper we improve EASG framework by replacing the most time-consuming part of the algorithm, solutions evaluation procedure, with a strategy evaluation neural network (SENN), to assess strategies from the current population. Figure~\ref{fig:overview} presents an overview of the proposed system. The multilayer perceptron SENN is first trained on historical data and then used to evaluate individuals in each generation.

\subsection{Strategy Evaluation Neural Network}
SENN is used to evaluate a given leader's strategy. Instead of exact calculation of the leader's payoff by finding an optimal follower's response (which is time-consuming and requires knowledge about follower's behaviour, preferences, or bounded rationality model) SENN approximates this value, taking the leader's mixed strategy as input
%, the output is one number - predicted expected leader's payoff while playing input strategy. 
The strategy is encoded in the following way. Each input node is feed with target coverage $c_s(t)$, i.e. a probability that at least one leader's unit is allocated to the target $t$ in time step $s$. Hence, the number of input neurons equals $nm$. 
%Neural network architecture is presented in Figure~\ref{fig:senn}. 
Neurons' outputs from the first layer are grouped by time step number - for each time step $\left \lceil{\frac{n}{4}}\right \rceil$ neurons are created - and then signals from all time steps are combined in the second hidden layer with $\left \lceil{\frac{n}{4}}\right \rceil$ neurons. A single output neuron returns the result.

\subsection{NESG}
SENN described in the previous section is incorporated into the EASG algorithm~\cite{ZychowskiMandziuk2021} leading to NESG (NeuroEvolutionary for Security Games) method.
%The resulting method is entitled Evolutionary Algorithm for Security Games with Strategy Evaluation Neural Network (NESG). 
NESG follows the general EASG protocol for solution finding. Each individual represents a possible solution, i.e. a mixed strategy of the leader. Initially, a population of individuals is generated, each of them representing a randomly selected pure strategy, i.e. leader's units are assigned to randomly chosen targets. Then, the following procedure is repeated until the limit for generation number is reached. A random subset of individuals is selected and divided into pairs. For each pair, a crossover operator is applied which is combines two strategies into one mixed strategy with halved probabilities. Then, mutation is performed on a randomly selected subset of the population. The mutation changes one pure strategy (in a mixed strategy represented by the mutated individual) by randomly changing the allocation of an arbitrary subset of leader's units. Afterward, the population is evaluated, i.e. for each individual the expected leader's payoff (assuming the strategy encoded by this individual is played) is calculated. Finally, the selection procedure picks individuals based on their evaluation and promotes them to the next generation. For further details please consult~\cite{ZychowskiMandziuk2021}.

EASG evaluation procedure iteratively checks all follower's strategies and selects the one with the highest follower's payoff~\cite{ZychowskiMandziuk2021}. The approach becomes inefficient for large games due to high number of strategies that need to be evaluated.
%to iterate and payoffs to calculate. 
In this paper the EASG evaluation procedure is replaced by SENN which estimates the expected leader's payoff.
%without directly calculating it and speed up computations.

\section{Experimental Setup}\label{sec:Setup}

\subsection{Benchmark games}
We tested NESG method on 90 randomly generated game instances which reflect real-world cybersecurity scenario described in Section~\ref{sec:Scenario}. For each number of time steps $m \in \{1,2,4\}$, 30 games were created with the number of targets $n=2^i$, $i \in \{2,\ldots,7\}$. Hence, there was exactly 5 games for each unique pair $(n,m)$. Payoffs $U_t^{L-}$ and $U_t^{F-}$ were real numbers drawn independently from interval $(-1;0)$, while $U_t^{L+}$ and $U_t^{F+}$ were sampled from $(0;1)$. The number of leader's units was drawn from interval $[\left \lfloor{\frac{n}{4m}}\right \rfloor ; \left \lceil{\frac{3n}{4m}}\right \rceil]$ (independently for each game instance ), i.e. at least $\frac{1}{4}$ and at most $\frac{3}{4}$ of the targets could be effectively protected.

\subsection{Bounded rationality models}
We considered three most popular BR models: anchoring theory, quantal response, and prospect theory.

\textbf{Anchoring theory (AT)}~\cite{tversky1974judgment} claims that humans tend to flatten probabilities of available options. In the decision process, they perceive a distribution of probabilities as being closer to the uniform distribution than it really is. 
Formally, according to AT, for any probability distribution over a finite set $X$ the observer perceives the probability of $q(x)$, $x\in X$ as $q'(x)=q(x)(1-\delta)+\delta/|X|$, where $0<\delta <1$ is a parameter of AT bias and $|X|$ is cardinality of $X$.

\textbf{Quantal response (QR)}~\cite{mckelvey1995quantal} assumes that humans choose a decision stochastically and the chance of selecting a non-optimal decision depends on the assigned payoff, i.e. the higher the payoff, the higher the chance for a decision to being chosen. According to QR the probability of making decision $x_i$ equals $q(x_i) = \frac{e^{\lambda u(x_i)}}{\sum_{x_k \in \Pi}e^{\lambda u(x_k)}}$ where $u(x)$ is expected payoff for decision $x$, $\Pi$ is a set of all possible decisions and $\lambda \geq 0$ is a parameter of QT bias. 

\begin{figure}[ht]
	\begin{center}
    \includegraphics[width=0.65\columnwidth]{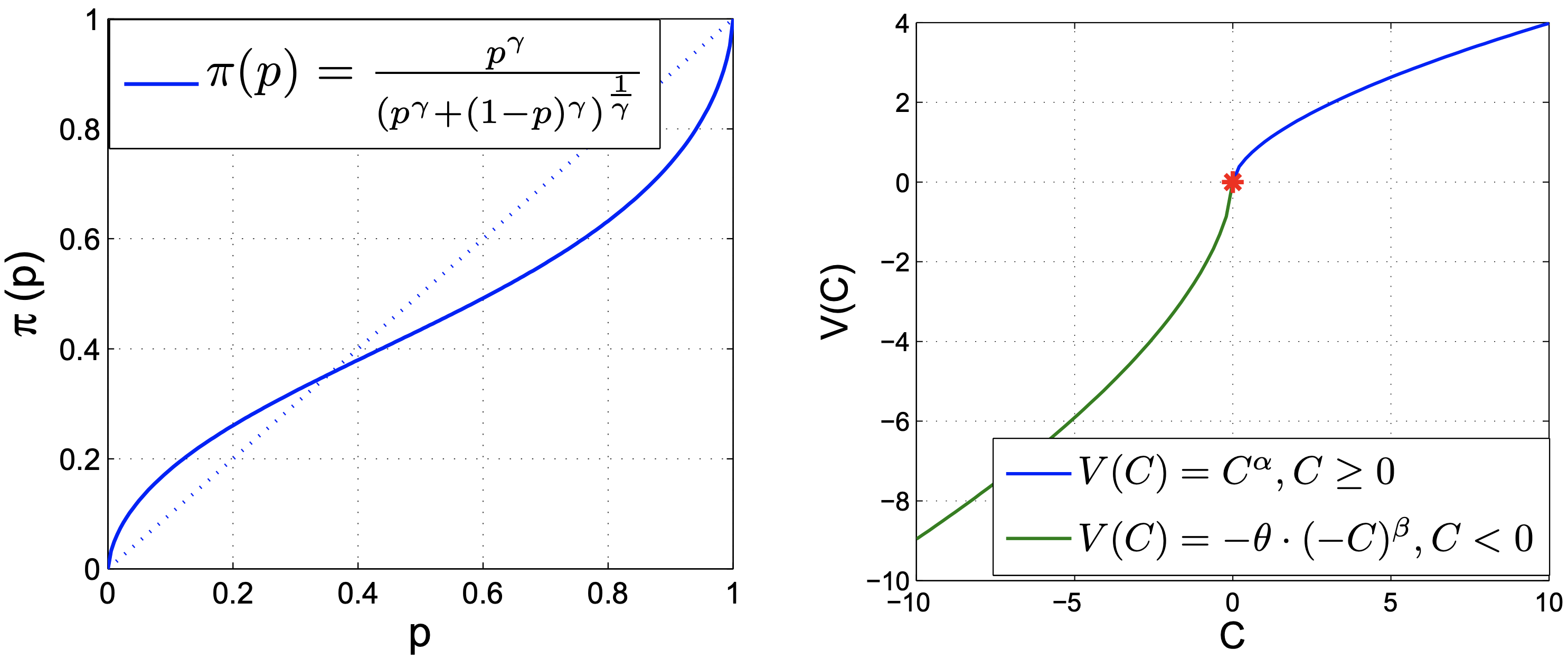}
    \caption{Prospect theory functions~\cite{yang2013improving}.}
    \label{fig:pt}
	\end{center}
\end{figure}

\textbf{Prospect theory (PT)}~\cite{kahneman2013prospect} bases on the concept that loss aversion and risk aversion are not symmetric. It was observed that instead of maximizing the expected payoff humans tend to maximize the \textit{prospect} which is defined as $\sum_i \pi(p_i)V(C_i)$ where $p_i$ is the actual probability of receiving payoff $C_i$. The weighting functions $\pi(\cdot)$ and $V(\cdot)$ describe people's perception of the probability and the outcome, resp. (see Figure~\ref{fig:pt}).

Please note that application of NESG is not limited to any particular BR model and the method can be used even in situations with no BR model defined. However, to make a comparison with other methods (which require formally defined BR model) we used 3 above mentioned popular models in our tests.
%with different characteristics - AT focuses on probabilities, QR on payoffs, PT on both of those aspects.

\subsection{Training data}
For each game, the historical data used to train SENN consisted of 5000 examples. Each of them was generated independently in the following way. 
The leader's mixed strategy was chosen randomly from all possible strategies with at most 5 pure strategies. Namely, the number of pure strategies $l$ was randomly chosen from 1 to 5. Then, $l$ pure strategies were generated by randomly allocating the leader's units and drawing the probability of this pure strategy from the unit interval. Finally, all probabilities were normalized so as to sum to 1.

It was assumed that the follower's decisions were consistent with the respective BR model (one of the three models described above) and, based on this assumption, the optimal follower's response was calculated (by brute-force iterating over all possible follower's strategies). This information allowed getting the exact leader's payoff (associated with the sampled strategy) which was saved as an expected SENN output for this training instance.

\subsection{Parameterization}
We followed EASG parameterization recommended in~\cite{ZychowskiMandziuk2021} with no additional parameter tuning: population size $=100$, number of generations $=1000$, mutation rate $=0.5$, crossover rate $=0.8$, selection in the form of a binary tournament with selection pressure $=0.9$ and elite size $=2$.

Please recall that SENN is a multilayer perceptron with $mn$, $m\left \lceil{\frac{n}{4}}\right \rceil$, $\left \lceil{\frac{n}{4}}\right \rceil$ and $1$ units in subsequent layers. SENN was trained with backpropagation with minibatch and Adam optimization. Hyperbolic tangent activation was used in the output node.
The following bias parameter values were assumed in BR models - AT: $\delta = 0.5$, QR: $\lambda = 0.8$, PT: $\gamma = 0.64, \theta = 2.25, \alpha = \beta = 0.88$.

\section{Experimental Results}\label{sec:Results}

\subsection{The quality of SENN training}
Table~\ref{tab:nnerror} shows SENN the average absolute difference between the network output and the true leader's payoff, calculated on the test set composed of 1000 samples generated in the same way as the training data (see Section~\ref{sec:Setup}). There are two main conclusions.
%raised from these results. 
Firstly, the network error increases with the number of targets and steps. This is an expected phenomenon since games become more complicated and the network has to process more data. Secondly, some differences between BR models can be observed. SENN predicts payoffs the more accurately for QR and AT than for PT. The probable reason is that PT nonlinearly affects both the perceived probabilities and the payoffs, whereas the other models change only one of these aspects: AT - probabilities, QR - payoffs.

\begin{table}[ht]
\begin{center}
\resizebox{0.8\textwidth}{!}{%
\begin{tabular}{c||c|c|c||c|c|c||c|c|c}
  & \multicolumn{3}{c||}{Anchoring Theory} & \multicolumn{3}{c||}{Quantal Response} & \multicolumn{3}{c}{Prospect Theory} \\ \hline
targets & 1 step  & 2 steps  & 3 steps & 1 step  & 2 steps  & 3 steps & 1 step  & 2 steps & 3 steps \\ \hline
4 & 0.006 & 0.006 & 0.006 & 0.004 & 0.005 & 0.005 & 0.010 & 0.010 & 0.010 \\ 
8 & 0.011 & 0.012 & 0.014 & 0.008 & 0.008 & 0.008 & 0.018 & 0.019 & 0.020 \\ 
16 & 0.024 & 0.026 & 0.028 & 0.019 & 0.021 & 0.022 & 0.046 & 0.049 & 0.051 \\ 
32 & 0.043 & 0.045 & 0.048 & 0.031 & 0.033 & 0.035 & 0.075 & 0.080 & 0.084 \\ 
64 & 0.080 & 0.081 & 0.086 & 0.064 & 0.065 & 0.069 & 0.132 & 0.142 & 0.145 \\ 
128  & 0.119 & 0.125 & 0.131 & 0.104 & 0.110 & 0.121 & 0.232 & 0.241 & 0.251
\end{tabular}
}
\end{center}
\caption{SENN error on test dataset. }
\label{tab:nnerror}
\end{table}

\subsection{Payoffs comparison}
In order to evaluate the practical efficacy of NESG, the method was compared with the following 5 methods: \textbf{C2016} - generates optimal (exact) solutions without considering follower's BR,
\textbf{EASG} - original EASG formulation (generates approximate solution with no BR model employed), \textbf{EASG\_BR} where BR$ \in \{$AT, QR, PT$\}$ - EASG method incorporating the respective BR model, i.e. the follower's response in the evaluation procedure is calculated assuming a given BR model.

For C2016 and EASG, first the leader's strategy was generated (without considering the follower's BR bias) and then the associated payoff was calculated under the assumption that the follower would actually not respond optimally but would follow a particular BR model.

Please note that it is not possible to incorporate BR models directly into MILP solutions (e.g. C2016) since BR models introduce \textit{nonlinear} modifications to payoffs or probabilities whose implementation in MILP would require using non-linear constraints.

\begin{table}[ht]
\centering
\begin{subtable}{1.0\textwidth}
\resizebox{1.0\textwidth}{!}{%
\begin{tabular}{c||c|c|c|c||c|c|c|c||c|c|c|c}

    \textbf{1 step\phantom{s}}    & \multicolumn{4}{c||}{Anchoring Theory}    & \multicolumn{4}{c||}{Quantal Response}    & \multicolumn{4}{c}{Prospect Theory}  \\ \hline
targets & \scriptsize{C2016} & \scriptsize{EASG}   & \scriptsize{EASG\_AT} & \scriptsize{NESG} & \scriptsize{C2016} & \scriptsize{EASG}   & \scriptsize{EASG\_QR} & \scriptsize{NESG} & \scriptsize{C2016} & \scriptsize{EASG}   & \scriptsize{EASG\_PT} & \scriptsize{NESG} \\ \cline{1-13}
4       & -0.470  & -0.472 & -0.468   & -0.469     & -0.406  & -0.408 & -0.404   & -0.405     & -0.419  & -0.420 & -0.417   & -0.418  \\
8       & -0.456  & -0.457 & -0.440   & -0.440     & -0.418  & -0.422 & -0.386   & -0.388     & -0.422  & -0.423 & -0.407   & -0.407  \\
16      & -0.387  & -0.391 & -0.371   & -0.371     & -0.377  & -0.378 & -0.336   & -0.338     & -0.329  & -0.335 & -0.315   & -0.318 \\
32      & -0.411  & -0.412 & -0.393   & -0.397     & -0.428  & -0.429 & -0.390   & -0.394     & -0.397  & -0.404 & -0.367   & -0.370 \\
64      & -0.579  & -0.586 & -0.567   & -0.568     & -0.582  & -0.584 & -0.536   & -0.537     & -0.560  & -0.564 & -0.483   & -0.486  \\
128     & -0.397  & -0.405 & -0.369   & -0.372     & -0.578  & -0.578 & -0.526   & -0.529     & -0.462  & -0.463 & -0.345   & -0.347 
\end{tabular}
}
%\caption{1 step games.}
\end{subtable}

\begin{subtable}{1.0\textwidth}
\resizebox{1.0\textwidth}{!}{%
\begin{tabular}{c||c|c|c|c||c|c|c|c||c|c|c|c}

    \textbf{2 steps}    & \multicolumn{4}{c||}{Anchoring Theory}    & \multicolumn{4}{c||}{Quantal Response}    & \multicolumn{4}{c}{Prospect Theory}  \\ \hline
targets & \scriptsize{C2016} & \scriptsize{EASG}   & \scriptsize{EASG\_AT} & \scriptsize{NESG} & \scriptsize{C2016} & \scriptsize{EASG}   & \scriptsize{EASG\_QR} & \scriptsize{NESG} & \scriptsize{C2016} & \scriptsize{EASG}   & \scriptsize{EASG\_PT} & \scriptsize{NESG} \\ \cline{1-13}
4 & -0.566  & -0.566 & -0.563   & -0.564     & -0.540  & -0.541 & -0.534   & -0.535     & -0.548  & -0.549 & -0.547   & -0.547     \\
8 & -0.568  & -0.572 & -0.553   & -0.555     & -0.526  & -0.528 & -0.510   & -0.512     & -0.556  & -0.556 & -0.517   & -0.518     \\
16& -0.327  & -0.331 & -0.314   & -0.317     & -0.326  & -0.331 & -0.301   & -0.302     & -0.326  & -0.331 & -0.291   & -0.294     \\
32& -0.499  & -0.500 & -0.475   & -0.479     & -0.487  & -0.487 & -0.435   & -0.435     & -0.501  & -0.502 & -0.454   & -0.457     \\
64& -0.457  & -0.463 & -0.427   & -0.427     & -0.421  & -0.424 & -0.403   & -0.408     & -0.466  & -0.471 & -0.407   & -0.410     \\
128 & -0.607  & -0.614 & -0.563   & -0.567     & -0.601  & -0.604 & -0.540   & -0.544     & -0.593  & -0.595 & -0.566   & -0.571
\end{tabular}
}
%\caption{2 steps games.}
\end{subtable}

\begin{subtable}{1.0\textwidth}
\resizebox{1.0\textwidth}{!}{%
\begin{tabular}{c||c|c|c|c||c|c|c|c||c|c|c|c}

    \textbf{4 steps}    & \multicolumn{4}{c||}{Anchoring Theory}    & \multicolumn{4}{c||}{Quantal Response}    & \multicolumn{4}{c}{Prospect Theory}  \\ \hline
targets & \scriptsize{C2016} & \scriptsize{EASG}   & \scriptsize{EASG\_AT} & \scriptsize{NESG} & \scriptsize{C2016} & \scriptsize{EASG}   & \scriptsize{EASG\_QR} & \scriptsize{NESG} & \scriptsize{C2016} & \scriptsize{EASG}   & \scriptsize{EASG\_PT} & \scriptsize{NESG} \\ \cline{1-13}
4 & -0.479   & -0.481 & -0.478 & -0.479 & -0.487   & -0.489 & -0.485 & -0.486 & -0.511   & -0.512 & -0.508 & -0.510 \\
8 & -0.497   & -0.500 & -0.466 & -0.467 & -0.509   & -0.513 & -0.455 & -0.456 & -0.517   & -0.519 & -0.496 & -0.499 \\
16   & -0.545   & -0.547 & -0.525 & -0.525 & -0.531   & -0.534 & -0.502 & -0.503 & -0.570   & -0.574 & -0.535 & -0.538 \\
32   & -0.478   & -0.484 & -0.460 & -0.464 & -0.500   & -0.505 & -0.468 & -0.470 & -0.525   & -0.531 & -0.492 & -0.496 \\
64   & -0.563   & -0.568 & -0.547 & -0.551 & -0.587   & -0.593 & -0.553 & -0.555 & -0.600   & -0.600 & -0.561 & -0.563 \\
128  & -0.531   & -0.536 & -0.493 & -0.497 & -0.545   & -0.549 & -0.503 & -0.505 & -0.553   & -0.555 & -0.512 & -0.512
\end{tabular}
}
%\caption{4 steps games.}
\end{subtable}

\caption{Average payoffs comparison for 1, 2 and 4 time steps games.}
\label{tab:results}
\end{table}

Table~\ref{tab:results} presents the average leader's payoffs for games with 1, 2 and 4 time steps. In all cases, NESG yielded better results than methods not considering (any) BR model (C2016 and EASG). The NESG advantage grows with the increasing number of targets. It can be concluded that when playing against not perfectly rational follower, it is better to use approximate NESG algorithm than playing the optimal strategy (generated by C2016) though without BR consideration.

For a given BR $\in \{$AT, QR, PT$\}$, the difference between EASG\_BR and NESG lays only in the population evaluation procedure. EASG\_BR computes the \textit{exact} follower's response which precisely employs a given BR model and then, based on that response, calculates the leader's payoff. NESG uses a neural network to \textit{estimate} the BR model of the follower.

Since EASG\_BR \textit{knows} the exact BR model and directly implements the BR function it should be treated as an \textit{oracle} model, whereas NESG is considered as its \textit{realistic} approximation relying on the past data in the network training process. Hence, NESG results presented in the tables are slightly worse than those of EASG\_BR.  Please note, however, that in a real-world scenario the leader, when calculating their strategy, \textit{is not aware} of the opponent's BR model. What is more the actual BR model of the follower would most likely differ from any of the 3 BR models considered in our experiments. In such cases using EASG\_BR with fixed BR model would be inefficient and the learning based NESG approach would gain significant advantage.

\subsection{Time scalability}
\begin{figure}[ht]
	\begin{center}
    \includegraphics[width=1.0\linewidth]{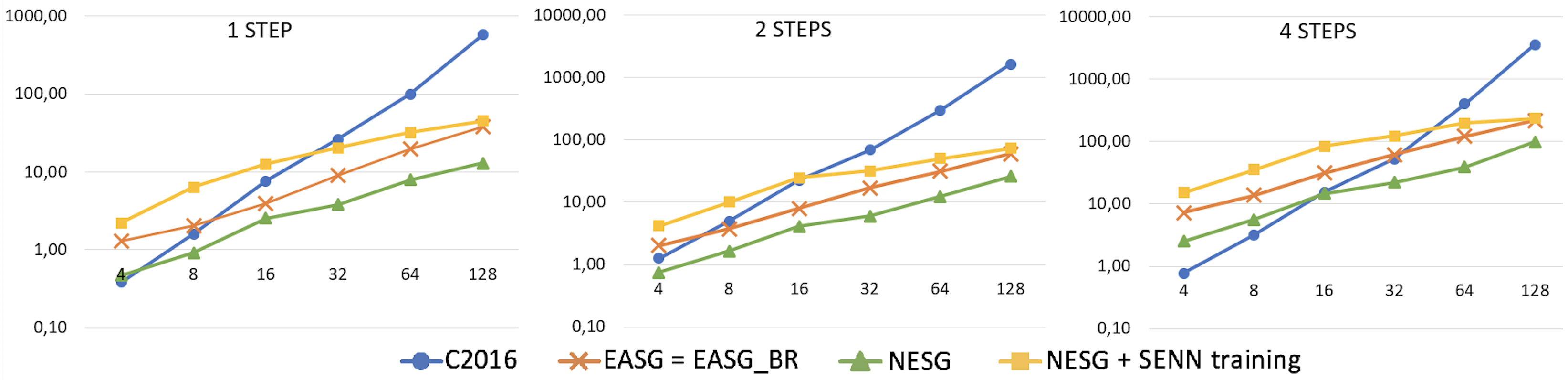}
    \caption{Comparison of NESG time scalability vs state-of-the-art methods. Please note a logarithmic scale on both axes.}
    \label{fig:time_scalability}
	\end{center}
\end{figure}

Figure~\ref{fig:time_scalability} shows time scalability of the proposed algorithm vs. other methods described in the previous section. To make fair comparison two variants of NESG computation time are presented: with and without SENN training time. Please note that typically the training procedure is performed once in a separate preprocessing stage and its time requirements do not affect the network inference process.

It can be concluded from the figure that C2016 (MILP based) scales visibly worse than evolutionary methods. Its computation time groves exponentially with respect to the number of targets. EASG, EASG\_BR and NESG scale much better in time and similar to each other. All 3 curves are roughly parallel and linear with respect to the number of targets. Computation times of EASG and EASG\_BR are the same because the difference between the methods lays only in the evaluation procedure. In EASG\_BR instead of getting the follower's payoff directly, BR model/function is imposed which requires just a couple of basic mathematical operations of meaningless cost. 

NESG computation time advantage over EASG and EASG\_BR stems from different ways of leader's strategy calculation. EASG / EASG\_BR require (1) iteration over all follower's strategies in order to find the best one, (2) marking it as follower's response and (3) computing the leader's payoff, which is time-consuming especially when the follower's strategy space is large. NESG gets an approximated leader's payoff directly from the neural network output.

\section{Conclusions}\label{sec:Conclusions}

This paper proposes a novel method for calculating leader's payoff in Stackelberg Security Games that uses strategy evaluation neural network (SENN). 
%can effectively approximate the leader's payoff for games playing against an opponent with bounded rationality. 
SENN is trained on historical data (results of previous games) with no explicit knowledge about the follower's payoff distribution or BR model, which well reflects real-world SG settings. In this paper, SENN is incorporated into evolutionary algorithm method (EASG), leading to an end-to-end SG solution (NESG), however, due to its generic nature, SENN can also be combined with other SG solution methods.
%It is the universal method for estimating the quality of a particular leader's strategy.
Experimental results on 90 benchmark games proven NESG efficacy and good time scalability. The method provides high quality result with low computation cost.

The main advantages of NESG are learning capabilities and knowledge-free nature.
%NESG  has strong potential for practical applications. 
Most of the existing algorithms require having full information about the attacker's payoff distribution, which is usually not possible in security management practical scenarios where defenders have only limited knowledge about their opponents. Furthermore, the existing methods typically assume perfect rationality of the attacker, which may not be the case in practice due to certain cognitive biases, wrong perception or imperfect information about the problem. NESG does not need to assume perfect rationality of the attacker and is able to infer the actual \textit{cognitive decision model / BR model} through learning. 

The proposed solution can potentially be applied to various real-life security situations e.g. homeland security, cybersecurity, banking fraud control or wildlife protection. One of particular applications in cybersecurity domain (deep packet inspections) is considered in this paper.

%Our future plan is to design a neural network for more complicated Security Games, e.g. games with uncertainty, limited opponent's observability, or underlying graph structure. Another possible research direction is to incorporate the proposed network into another SG methods.

\section*{Acknowledgment}
The project was funded by POB Research Centre Cybersecurity and Data Science of Warsaw University of Technology within the Excellence Initiative Program - Research University (ID-UB).

\bibliographystyle{unsrt}  
\bibliography{easgnn}  % put name of your .bib file here

\end{document}